\documentclass[preprint,aps,prd,amsmath,amssymb,floatfix]{revtex4}
\pdfoutput=1
\usepackage{graphicx}
\usepackage{color}
\newcommand{\be}{\begin{equation}}
\newcommand{\ee}{\end{equation}}
\newcommand{\bea}{\begin{eqnarray}}
\newcommand{\eea}{\end{eqnarray}}
\newcommand{\beq}{\begin{eqnarray}}
\newcommand{\eeq}{\end{eqnarray}}

\newcommand{\bmp}{\noindent\begin{minipage}{16cm}}
\newcommand{\emp}{\end{minipage}\vskip 7mm} 

\usepackage{graphicx}
\usepackage{dcolumn}
\usepackage{bm}
\usepackage{amsmath}
\usepackage{amsfonts}
\usepackage{amssymb}
\usepackage{bbm}
\usepackage{subfigure}
\usepackage{pxfonts}
\usepackage{slashed}

\definecolor{rossoCP3}{cmyk}{0,.88,.77,.40}

\usepackage{fancyhdr}
\usepackage{ulem}

\def\drawbox#1#2{\hrule height#2pt
        \hbox{\vrule width#2pt height#1pt \kern#1pt
              \vrule width#2pt}
              \hrule height#2pt}

\def\Asym#1#2{\vcenter{\vbox{\drawbox{#1}{#2}
              \kern-#2pt 
              \drawbox{#1}{#2}}}}

\begin{document}
\title{\Large  \color{rossoCP3}   Extreme Technicolor \\ \&  \\ The Walking Critical Temperature }
\author{Matti J\"arvinen$^{{\color{rossoCP3}{\spadesuit}}{\color{rossoCP3}{\varheartsuit}}}$}
\email{mjarvine@physics.uoc.gr} 
\author{Francesco Sannino$^{\color{rossoCP3}{\varheartsuit}}$}
\email{sannino@cp3.sdu.dk} 
\affiliation{
$^{\color{rossoCP3}{\spadesuit}}$ Crete Center for Theoretical Physics, Department of Physics, University of Crete, 71003 Heraklion, Greece \\
$^{\color{rossoCP3}{\varheartsuit}}${ CP}$^{ \bf 3}${-Origins}, 
Campusvej 55, DK-5230 Odense M, Denmark. }
\begin{flushright}
{\it CP$^3$- Origins-2010-39}\\
{\it CCTP-2010-12}
\end{flushright}
\begin{abstract}
We map the phase diagram of gauge theories of fundamental interactions in the flavor-temperature plane using chiral perturbation theory to estimate the relation between the pion decaying constant and the critical temperature above which chiral symmetry is restored. We then investigate the impact of our results on models of dynamical electroweak symmetry breaking and  therefore on the electroweak early universe phase transition. 
\end{abstract}

\maketitle
 \thispagestyle{fancy}
\section {Extreme Technicolor}

Arguably technicolor extensions of the standard model \cite{Weinberg:1979bn,Susskind:1978ms} constitute a natural mechanism able to explain and drive the breaking of the electroweak symmetry (see the recent review \cite{Sannino:2009za}).

As for ordinary Quantum Chromodynamics we can consider technicolor models in extreme conditions of temperature and matter density \cite{Braun:2006jd, Kikukawa:2007zk,Cline:2008hr,Jarvinen:2009pk,Jarvinen:2009wr,Braun:2009ns}. These regimes are particularly interesting for the early universe dynamics and phenomenology \cite{Cline:2008hr,Jarvinen:2009pk,Jarvinen:2009wr}. {}For example the technicolor temperature driven chiral phase transition is directly mapped into the electroweak phase transition. Applications to the detection of gravitational waves is very interesting and can potentially distinguish different scenarios of dynamical electroweak symmetry breaking \cite{Jarvinen:2009mh}. 

Adding a nonzero matter density is also very interesting for investigating asymmetric type dark matter emerging naturally as a technicolor interacting massive particle. The early models dealt with scaled up version of the ordinary baryons, i.e. technibaryon \cite{Nussinov:1985xr,Chivukula:1989qb,Barr:1990ca}.  Recent models of technicolor  \cite{Sannino:2009za} led to the introduction of new types  of dark matter candidates \cite{Gudnason:2006ug,Gudnason:2006yj,Kouvaris:2007iq,Kouvaris:2008hc,Foadi:2008qv,Kainulainen:2006wq,Khlopov:2007ic,Khlopov:2008ty,Ryttov:2008xe,Frandsen:2009mi,Belyaev:2010kp}. Introducing nonzero chemical potentials allows us to investigate the dark matter composition and distribution in the universe. An intriguing point is that for the recent relevant models of technicolor \cite{Sannino:2009za}  the action remains positive upon introduction of the chemical potential and therefore the nonperturbative dynamics can be studied using first principle lattice computations. This is so since in these models the {\it sign problem} is absent. An initial study of the nonzero matter density and temperature phase diagram, highly relevant for the new dark matter candidates, has been presented in \cite{Sannino:2004ix}. Here it was also shown that there is an intriguing relation between the matter density driven chiral phase transition and the deconfining one yielding a very rich phase diagram as function of the temperature and matter density.

In this work we start mapping the phase diagram of gauge theories of fundamental interactions in the flavor-temperature plane using chiral perturbation theory to estimate the relation between the pion decaying constant and the critical temperature above which chiral symmetry is restored. We then investigate the impact of our results on models of dynamical electroweak symmetry breaking and  therefore on the electroweak early universe phase transition.

\section{Chiral Lagrangian set up}
If the number of flavors is sufficiently low we expect chiral symmetry to break for any gauge theory of fundamental interactions we consider here. It is, hence, always possible to construct a chiral Lagrangian in terms of the Goldstone bosons of the theory and compute the temperature dependence of the chiral condensate \cite{Gasser:1984gg,Kogan:1984nb,Gasser:1986vb,Gerber:1988tt,Leutwyler:1992yt,Smilga:1994tb,Bijnens:2009qm}. At the leading order in the chiral expansion the only scale in the problem is the pion decay constant.

\subsection{Single Matter Field Representation}
First, we will discuss the theories with fermions transforming under a single representation of the technicolor gauge group. For a theory with $2N_f$ Weyl fermions, the chiral symmetry $G$ is either $SU(N_f)_L \times SU(N_f)_R$ or extended to $SU(2N_f)$, if the representation is complex or (pseudo)real, respectively. We expect that the chiral symmetry breaks spontaneously to its maximal diagonal subgroup $H$. This leads to the following chiral symmetry breaking patterns $G \to H$:
\begin{itemize}
 \item $SU(N_f)_L \times SU(N_f)_R \to SU(N_f)_V$ for complex representations,
 \item $SU(2N_f) \to SO(2N_f)$ for real representations,
 \item $SU(2N_f) \to Sp(2N_f)$ for pseudoreal representations.
\end{itemize}
Examples of (minimal) technicolor theories \cite{Sannino:2009za,Sannino:2004qp,Dietrich:2005jn,Dietrich:2006cm,Ryttov:2008xe,Sannino:2009aw,Frandsen:2009mi} with some of the above chiral symmetry breaking structures are:
\begin{itemize}
 \item Minimal Walking Technicolor (MWT), which has $N_f=2$ in the adjoint of $SU(2)_{TC}$ with expected pattern $SU(4)\to SO(4)$.
 \item Next-to-Minimal Walking Technicolor (NMWT), which has $N_f=2$ in the symmetric representation of $SU(3)_{TC}$ with the pattern $SU(2)_L \times SU(2)_R \to SU(2)$. 
\end{itemize}
We shall construct the nonlinear chiral Lagrangians following Refs. \cite{Ryttov:2008xe,Gudnason:2006ug} (other kinds of effective theories for minimal walking techicolor models have been considered in \cite{Sannino:2009za,Ryttov:2008xe,Gudnason:2006ug,Appelquist:1999dq,Foadi:2007ue,Belyaev:2008yj,Foadi:2008xj}). In a basis formed by the  Weyl fermions $\left(Q^L_1,\ldots ,Q^L_{N_f},i\sigma_2 \left(Q^{R}_1\right)^*,\ldots,i\sigma_2\left(Q^{R}_{N_f}\right)^*\right)$, the chiral condensate is proportional to a $2 N_f\times 2N_f$ matrix which we denote by:
\be
 E = \left(\begin{array}{cc}
      0 & \mathbf{1} \\
      \pm \mathbf{1} & 0
     \end{array}\right) \ ,
\ee
with the plus sign for complex or real representations, and minus sign for pseudoreal representations. Here the subblocks are $N_f\times N_f$ matrices. We denote the unbroken generators by $S^a$. They leave $E$ invariant: $(S^a)^T E + E S^a = 0$. The broken generators are denoted by $X^i$. Explicit expressions for the generators can be found in \cite{Appelquist:1999dq,Gudnason:2006ug}. Let us define the element $\mathcal{V}$ of the coset space 
\be
 {\cal V} = \exp\left(\frac{i}{F_\pi}\sum_i \Pi_i X_i\right)E \ ,
\ee
where $F_\pi$ is the Goldstone boson decay constant. It transforms non-linearly
\begin{eqnarray} \label{trans}
\mathcal{V}(\xi) \rightarrow g \mathcal{V}(\xi) h^{\dagger}(\xi, g)
\end{eqnarray}
where $g$ is an element of $G$ and $h$ is an element of $H$. We can embed the electroweak gauge group in $SU(4)$ as done in \cite{Ryttov:2008xe,Appelquist:1999dq}.
It is appropriate to introduce the Hermitian, algebra valued, Maurer-Cartan one-form
\begin{eqnarray}
\omega_{\mu} = i \mathcal{V}^{\dagger} \partial_{\mu} \mathcal{V}
\end{eqnarray}

From the above transformation properties of $\mathcal{V}$ it is clear that $\omega_{\mu}$ transforms as
\begin{eqnarray}
\omega_{\mu} \rightarrow h(\xi, g) \omega_{\mu} h^{\dagger}(\xi, g) + h(\xi, g) \partial_{\mu} h^{\dagger}(\xi, g) \ .
\end{eqnarray}
With $\omega_{\mu}$ taking values in the algebra of $G$ we can decompose it into a part $\omega_{\mu}^{\parallel}$ parallel to $H$ and a part $\omega_{\mu}^{\perp}$ orthogonal to $H$
\begin{eqnarray}
\omega_{\mu}^{\parallel} = 2S^a \text{Tr}\left[ S^a \omega_{\mu} \right] \ , \qquad \omega_{\mu}^{\perp} = 2X^i \text{Tr} \left[ X^i \omega_{\mu} \right] \ .
\end{eqnarray}
Then $\omega_{\mu}^{\parallel}$ is an element of the algebra of $H$ while $\omega_{\mu}^{\perp}$  of $G/H$. We find the following transformation properties:
\begin{eqnarray}
\omega_{\mu}^{\parallel} \rightarrow h(\xi, g) \omega_{\mu}^{\parallel} h^{\dagger}(\xi, g) + h(\xi, g) \partial_{\mu} h^{\dagger}(\xi, g) \ , \qquad \omega_{\mu}^{\perp} \rightarrow h(\xi, g) \omega_{\mu}^{\perp} h^{\dagger}(\xi, g) \ .
\end{eqnarray}
{}To probe the chiral dynamics of a theory we introduces a small, with respect to the dynamically generated scale associated to the chiral condensate, democratic mass matrix $\propto E$ for the fermions at the underlying theory level. Since the mass term introduces an explicit breaking of the chiral symmetry the pions become massive. The democratic choice of the mass leads to equal masses to the Goldstone bosons. At the effective Lagrangian level we have:  
\begin{eqnarray}
\mathcal{L} =  F_\pi^2 \text{Tr} \left[ \omega_{\mu}^{\perp} \omega^{\mu\perp}  \right]  - \frac{m_\pi^2F_\pi^2}{4}\text{Tr} \left[E^\dagger {\cal V}^T E {\cal V}+ \text{h.c.}\right]\ .
\end{eqnarray}

\subsection{Multiple Representations: The Ultra Minimal Technicolor example}

Lagrangians for technicolor models featuring matter transforming under several distinct representations of the underlying gauge group may be constructed in a similar fashion. The minimal walking model known as Ultra Minimal Technicolor (UMT) is one of the phenomenologically relevant examples. It contains two Dirac fermions transforming according to the fundamental representation and two Weyl fermions belonging to the adjoint representations of the technicolor gauge group $SU(2)_{TC}$. Its nonlinear Lagrangian was constructed in \cite{Ryttov:2008xe}. UMT has one anomaly free $U(1)$ symmetry.  We expect the nonperturbative dynamics to spontaneously break the global $SU(4)\times SU(2) \times U(1)$ symmetry to $Sp(4) \times SO(2)\times Z_2$ via the formation of the two distinct condensates. Let us order the broken generators such that $\left\{X_1,\ldots,X_5\right\}$, $\left\{X_6,X_7\right\}$, and $X_8$ are part of the generators of $SU(4)$, $SU(2)$, and $U(1)$, respectively (see \cite{Ryttov:2008xe} for an explicit realization). We define
\begin{eqnarray}
E= \left( \begin{array}{cc}
E_4 & \\
  & E_2
\end{array}\right) \ ,
\end{eqnarray}
where
\begin{equation}
 E_4= \left( \begin{array}{cc}
0 & \mathbf{1}\\
-\mathbf{1}  & 0
\end{array}\right) \ ; \qquad 
 E_2= \left( \begin{array}{cc}
0 & 1\\
1  & 0
\end{array}\right) \ .
\end{equation}
The element $\cal V$ can be parameterized by
\begin{eqnarray}
\mathcal{V}(\xi) = \exp \left( i \xi^iX^i \right) E \ ,
\end{eqnarray}
where
\begin{eqnarray}
 \xi^i X^i = \sum_{i=1}^{5} \frac{\Pi^iX^i}{F_{\pi}} + \sum_{i=6}^{7} \frac{\Pi^iX^i}{\tilde{F}_{\pi}} + \frac{\Pi^8 X^8}{\hat{F}_{\pi}} \ ,
\end{eqnarray}
and $F_{\pi}, \tilde{F}_{\pi}$ and $\hat{F}_{\pi}$ are the related Goldstone boson decay constants.  We introduce independent mass terms for the fundamental and adjoint techniquarks. We follow the procedure outlined above to construct nonlinear Lagrangians and by noting that the generator $X^8$ corresponding to the broken $U(1)$ is not traceless we deduce:
\begin{eqnarray}
\mathcal{L} = \text{Tr} \left[a \omega_{\mu}^{\perp} \omega^{\mu\perp}  \right] + b \text{Tr} \left[ \omega_{\mu}^{\perp} \right] \text{Tr} \left[ \omega^{\mu \perp} \right] - \text{Tr} \left[ c E^\dagger {\cal V}^T E {\cal V}+ \text{h.c.}\right]\ ,
\end{eqnarray}
The coefficients $a=\text{diag}\left(F_{\pi}^2, F_{\pi}^2, F_{\pi}^2, F_{\pi}^2, \tilde{F}_{\pi}^2, \tilde{F}_{\pi}^2 \right)$ and $b=\frac{\hat{F}_{\pi}^2}{2} - \frac{4F_{\pi}^2}{9} - \frac{\tilde{F}_{\pi}^2}{18}$ are chosen such that the kinetic term is canonically normalized:
\begin{eqnarray}
\mathcal{L} = \frac{1}{2} \sum_{i=1}^8 \partial_{\mu} \Pi^i \partial^{\mu} \Pi^i + \ldots \ .
\end{eqnarray}
The remaining  coefficient is $c=\text{diag}\left(m_{\pi}^2F_{\pi}^2/4,m_{\pi}^2F_{\pi}^2/4,m_{\pi}^2F_{\pi}^2/4,m_{\pi}^2F_{\pi}^2/4,\tilde{m}_{\pi}^2 \tilde{F}_{\pi}^2/4, \tilde{m}_{\pi}^2 \tilde{F}_{\pi}^2/4 \right)$, where $m_{\pi}$ ($\tilde{m}_{\pi}$) is the mass of the pions emerging due to the presence of the fermions in the fundamental (adjoint) representation. The mass of the eighth pion, related to the breaking of the anomaly-free $U(1)$, is given by
\be
 \hat m_\pi^2 = \frac{1}{9\hat F_\pi^2}\left[8 F_\pi^2 m_\pi^2 + \tilde m_\pi^2 \tilde F_\pi^2 \right] \ .
\ee

\section{Raising the Temperature}
\label{rt}
Next we shall calculate the pressure and the chiral condensates at finite temperature up to two loops. The pressure can be calculated following \cite{Gerber:1988tt}. At one loop, it equals the Bose contribution from a pion gas. Fortunately the two-loop finite temperature results are independent on  higher derivative terms of the effective zero temperature Lagrangian. To this order, therefore, we can be predictive. Beyond two loops one needs the coefficients of higher derivative terms that cannot yet be derived from experiments.  Lattice simulations, in the future, can provide a systematic study of the chiral properties of these theories. 

We summarize below the two-loop results for the pressure for the different patterns of chiral symmetry breaking introduced above:  
\bea
 p(T) &=&  \frac{N_f^2-1}{2}g_0(m_\pi,T)- \frac{(N_f^2-1)m_\pi^2}{8N_fF_\pi^2}g_1(m_\pi,T)^2 \ ; \qquad \text{complex} \\
 p(T) &=&  \frac{(N_f+1)(2N_f-1)}{2}g_0(m_\pi,T) + \frac{(N_f^2-1)(2N_f-1)m_\pi^2}{8N_fF_\pi^2}g_1(m_\pi,T)^2 \ ; \qquad \text{real} \\
 p(T) &=&  \frac{(N_f-1)(2N_f+1)}{2}g_0(m_\pi,T) - \frac{(N_f^2-1)(2N_f+1)m_\pi^2}{8N_fF_\pi^2}g_1(m_\pi,T)^2 \ ; ~~\text{pseudoreal}  \\
 p(T) &=&  \frac{5}{2}g_0(m_\pi,T) + g_0(\tilde m_\pi,T) + \frac{1}{2}g_0(\hat m_\pi,T) \\
 && -\frac{45}{48} \frac{m_\pi^2}{F_\pi^2}g_1(m_\pi,T)^2 +\frac{5}{9} \frac{m_\pi^2}{\hat F_\pi^2}g_1(m_\pi,T)g_1(\hat m_\pi,T) +\frac{1}{18} \frac{\tilde m_\pi^2}{\hat F_\pi^2}g_1(\tilde m_\pi,T)g_1(\hat m_\pi,T) \\
 &&+\left(\frac{4}{81}m_\pi^2F_\pi^2 + \frac{1}{648}\tilde m_\pi^2\tilde F_\pi^2\right) \frac{g_1(\hat m_\pi,T)^2}{\hat F_\pi^4} \ ; \qquad \text{UMT} \ ,
\eea
where
\be
 g_r(m_\pi,T)= 2\int_0^\infty \frac{d\lambda {\lambda^{r-1}}}{(4\pi\lambda)^2}e^{-\lambda m_\pi^2}\sum_{n=1}^\infty e^{-\frac{n^2}{4\lambda T^2}} \ .
\ee

When we have matter in a single representation, the temperature dependent techniquark condensate is found by taking the derivative of the pressure with respect to the quark mass. We find
\be
 \frac{\left\langle\bar Q Q \right\rangle}{\left\langle\bar Q Q \right\rangle_{T=0}} = 1 + \frac{c}{N_f F_\pi^2}\frac{\partial p(T)}{\partial m_\pi^2}
\ee
where the coefficient $c$ approaches unity as $m_\pi$ goes to zero, and we needed also the zero temperature GMOR relation:
\begin{equation}
 \left\langle\bar Q Q\right\rangle_{T=0} = -\frac{ N_f F_{\pi}^2 m_\pi^2}{m} \ .
\end{equation}
For the first three cases corresponding to a single representation we have: %
\bea
 \frac{\left\langle\bar Q Q \right\rangle}{\left\langle\bar Q Q \right\rangle_{T=0}} &=& 1 - \frac{(N_f^2-1)T^2}{24 N_f F_\pi^2} - \frac{(N_f^2-1)T^4}{1152 N_f^2 F_\pi^4}\ ; \qquad \text{complex} \\
 &=& 1 - \frac{(N_f+1)(2N_f-1)T^2}{24 N_f F_\pi^2} + \frac{(N_f^2-1)(2N_f-1)T^4}{1152 N_f^2 F_\pi^4}\ ; \qquad \text{real} \\
 &=& 1 - \frac{(N_f-1)(2N_f+1)T^2}{24 N_f F_\pi^2} - \frac{(N_f^2-1)(2N_f+1)T^4}{1152 N_f^2 F_\pi^4}\ ; \qquad \text{pseudoreal} 
\eea
as $m_\pi$ is set to zero.
For UMT, we get for the condensate of the fundamental quarks
\bea
 \frac{\left\langle\bar Q Q \right\rangle}{\left\langle\bar Q Q \right\rangle_{T=0}} &=& 1 + \frac{c}{2 F_\pi^2}\frac{\partial p(T)}{\partial m_\pi^2} + \frac{4c}{9 \hat F_\pi^2}\frac{\partial p(T)}{\partial \hat m_\pi^2} \\
 &\stackrel{m_\pi, \hat{m}_{\pi} = 0}{=}& 1 -\left(\frac{5}{48}+ \frac{1}{54}\frac{F_\pi^2}{\hat F_\pi^2}\right)\frac{T^2}{F_\pi^2}-\left(\frac{5}{1536}-\frac{5}{2592}\frac{F_\pi^2}{\hat F_\pi^2}-\frac{1}{5832}\frac{F_\pi^4}{\hat F_\pi^4}\right)\frac{T^4}{F_\pi^4} \ .
\eea

\begin{figure}[t]
{\includegraphics[height=6.5cm]{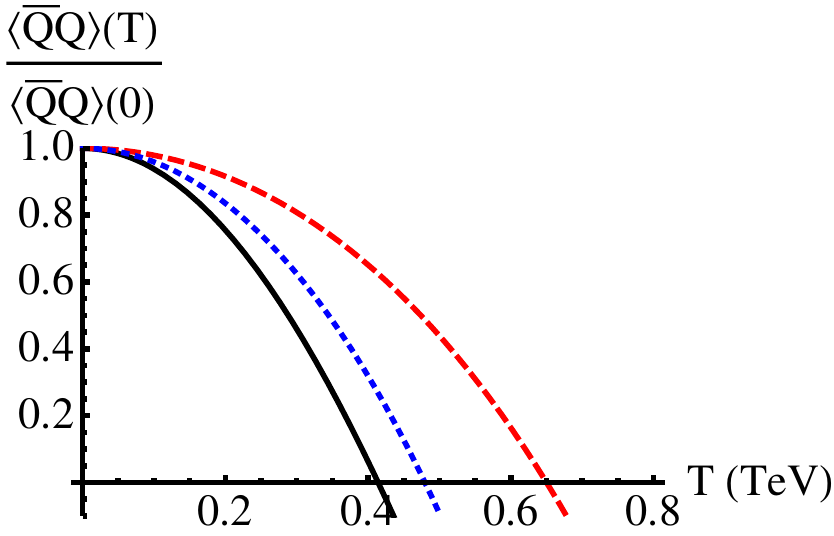}\hspace{0.8cm}}
\caption{The temperature dependence of the two-loop chiral condensate for MWT (black solid curve), NMWT (red dashed curve), and UMT (blue dotted curve)}\label{fig:condplot}
\end{figure}

Let us plot the result for some phenomenologically relevant technicolor models. We fix the condensate to the electroweak scale via the relation 
\begin{equation}
v_{\rm EW} = \sqrt{N_{fg}} \,\,\,F_\pi \ , \quad {\rm with } \quad v_{\rm EW} \simeq 246~{\rm GeV} \ .
\end{equation}
 $N_{fg}$ is the number of flavors gauged under the electroweak symmetry. In Figure~\ref{fig:condplot} we plot the condensates as function of the temperature for the MWT, NMWT, and UMT models. {}For (N)MWT we fixed $F_\pi$ to the electroweak scale setting $N_{fg}=2$. {}For the UMT sector only the techniquarks transforming according to the fundamental representation of the $SU(2)$ technicolor gauge group are gauged under the electroweak symmetry and therefore the decay constant $F_\pi$ of the fundamental sector was fixed to electroweak with $N_{fg}=2$. We assumed $F_\pi=\hat F_\pi$ since with two technicolors the fundamental and adjoint representations are expected to lead to similar screening effects.   
 
 The temperature effects, to the order we have computed, tend to reduce the size of the condensate with respect to the zero temperature value. To estimate the critical temperature above which chiral symmetry is expected to restore we set the  temperature dependent condensate to zero and read off the associated critical temperature. This procedure is not expected to yield a precise estimate of the critical temperature but should capture the essential features related to the chiral symmetry breaking pattern on the underlying gauge theory. 
 
Therefore, by solving for the temperature where the condensate melts we deduce $T_c=414$, 650, and $480$~GeV for MWT, NMWT, and UMT respectively. 

It is interesting also to investigate the general case in which we gauge under the electroweak symmetry all doublets: $2 N_D = N_f=N_{fg}$. The critical temperature can than be approximated analytically using the one-loop result once fixed $F_\pi$ to the electroweak scale and reads: 
\bea
 T_c & \simeq & \frac{\sqrt{24 N_f}}{\sqrt{N_f^2-1}}F_\pi = \frac{\sqrt{24}}{\sqrt{N_f^2-1}}\,v_{\rm EW} \ ; \qquad \text{complex} \\
           & \simeq & \frac{\sqrt{24 N_f}}{\sqrt{(N_f+1)(2N_f-1)}}F_\pi = \frac{\sqrt{24}}{\sqrt{(N_f+1)(2N_f-1)}}\, v_{\rm EW} \ ;\qquad \text{real} \\
           & \simeq & \frac{\sqrt{24 N_f}}{\sqrt{(N_f-1)(2N_f+1)}}F_\pi = \frac{\sqrt{24}}{\sqrt{(N_f-1)(2N_f+1)}}\, v_{\rm EW} \ ;\qquad \text{pseudoreal}
  \label{onetemperature}
\eea
It is clear from the formulae just above that the critical temperature is, for large $N_f$, inversely proportional to the number of flavors. 

\section{Walking Critical Temperature}

In the previous sections we studied the dependence of the condensate as function of the temperature for different technicolor models at fixed number of flavors and matter representation. We have also fixed the pion decay constant in order to reproduce the correct W gauge boson mass. 

We now deduce the dependence of the critical temperature on the number of flavors for a given gauge theory as we approach the critical number of flavors above which the theory is expected, at zero temperature, to develop a nonzero infrared fixed point. 
To do so we normalize the {\it walking} temperature $T_c [N_f]$ to $T_c[\bar{N}_f=2]$ for the complex and pseudoreal representations. We used the two-loop finite temperature result given above to deduce the intrinsic dependence on the zero temperature pion decay constant. To get an idea of the $N_f$ dependence we show the one-loop result which fits in one line:   
\begin{eqnarray}
\frac{T_c [N_f]}{T_c[2]} &=&  
{\frac{\sqrt{3 \, N_f}}{\sqrt{2(N_f^2 -1)} }} \frac{F_{\pi}[N_f]}{F_{\pi}[2]} \ , \qquad {\rm complex} \ , \\
   \frac{T_c [N_f]}{T_c[2]} &=&  
{\frac{\sqrt{5 \, N_f}}{\sqrt{2(N_f -1)(2N_f+1)} }} \frac{F_{\pi}[N_f]}{F_{\pi}[2]} \ , \qquad {\rm pseudoreal} \ . \\
\end{eqnarray}
From the expressions above it is clear that one needs the explicit dependence of the pion decay constant as function of the number flavors. 
Near the lower end of the conformal window one can estimate such a dependence using the Schwinger-Dyson results in the rainbow approximation \cite{Appelquist:1998xf,Sannino:2004qp,Dietrich:2006cm}. We start with the well known relation: 
\begin{equation}
2\pi \,F_{\pi}[N_f] \approx \sqrt{d\left[r\right]} \,  \Sigma_0 [N_f]
\end{equation} 
with $d\left[r\right]$ the dimension of the representation of a technifermion species with respect to the underlying gauge group and the dynamical mass:
\begin{equation}
\Sigma_0 [N_f]= \Lambda_c [N_f] \exp \left(-\frac{1}{\sqrt{\frac{\alpha_{\ast}[N_f]}{\alpha_c} -1}}\right) \ . 
\label{sigma}
\end{equation}
Since we are comparing a given gauge theory for different number of flavors, in this section, we do not indicate the explicit dependence on the fermion representation and number of colors which must, however, be taken properly into account when computing the walking temperature for each theory. 

\begin{figure}[t]
{\includegraphics[height=5.5cm]{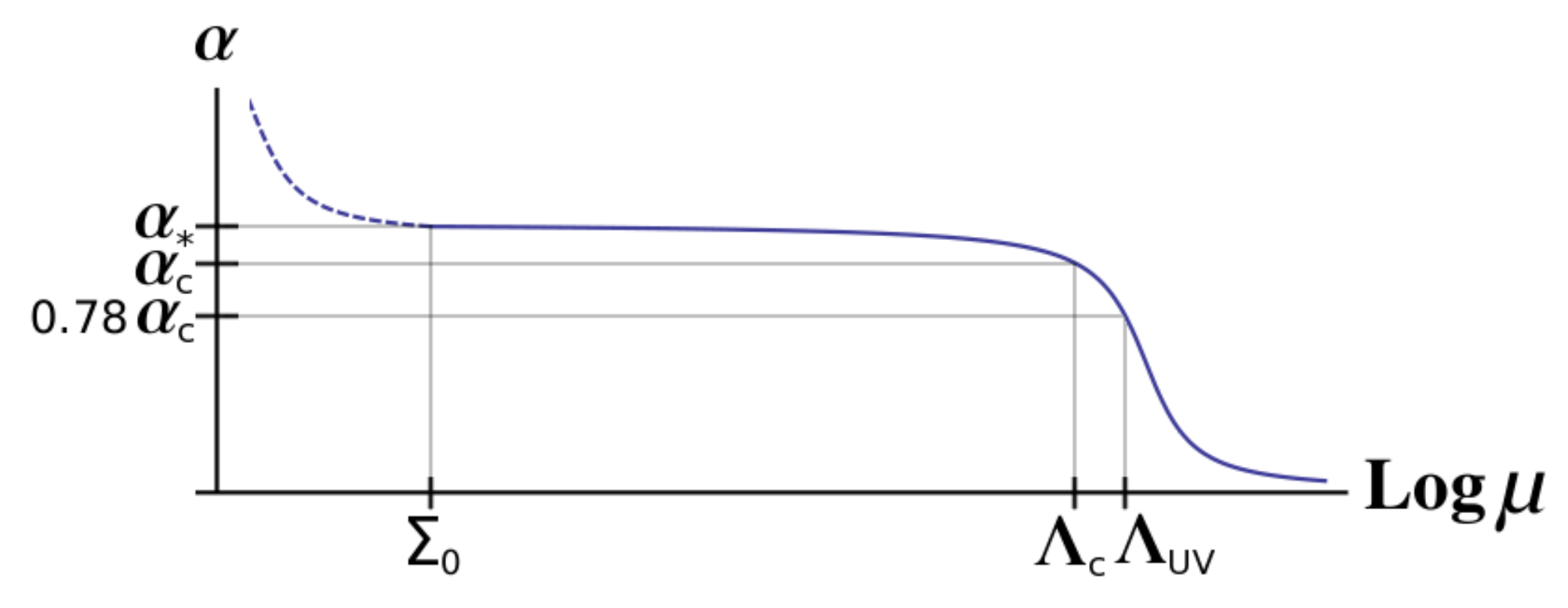}\hspace{0.8cm}}
\caption{The definitions of various scales and couplings in the walking region, with $\alpha_c$ close to $\alpha_{\ast}$. }\label{fig:hierarchy}
\end{figure}

Here $\alpha_{\ast}[N_f]$ is the infrared fixed coupling at two loops:
\begin{equation}
{\alpha_{\ast}[N_f]} = - \frac{\beta_0}{\beta_1} \ ,  
\end{equation}
with the following definition of the two-loop beta function:
\begin{equation}
\beta(\alpha) = -{\beta_0} \alpha^2 - {\beta_1} \alpha^3 \ .
\end{equation}
We also have:
\begin{equation}
\alpha_c = \frac{\pi}{3C_2[r]} \ ,
\end{equation}
which is the critical value of the coupling below which one expects the underlying theory to develop a nontrivial infrared fixed point and $C_2[r]$ is the quadratic casimir. The number of flavors is assumed to be sufficiently large that the underlying gauge theory breaks dynamically chiral symmetry. Since we are approaching the conformal window from the broken side $\alpha_{\ast}$ is the would be infrared fixed point if chiral symmetry would not be broken according to the two-loop beta function. $\Lambda_c$, following \cite{Appelquist:1998xf}, is defined to be the energy scale at which the coupling, as function of the renormalization scale, crosses $\alpha_c$. Another relevant scale is the one above which the theory is expected to run as in QCD. This is the technical scale $\Lambda_{UV}$ usually defined such that the coupling evaluated at this scale is about $0.78 \, \alpha_c$ \cite{Appelquist:1997fp}.  The above is the standard rainbow approximation scenario and the hierarchy of the scales is represented in Fig.~\ref{fig:hierarchy}.

Because of the existence of a lower number of flavors changing the sign of the second coefficient of the two-loop beta function the above approximation of $F_{\pi}$ as function of the number of flavors does not automatically extend to the lowest possible value of the number of flavors. When the second coefficient vanishes the exponential in \eqref{sigma} approaches unity. Below the associated value of the number flavors we simply take $\Sigma_0[N_f] = \Lambda_c [N_f]$.

\begin{figure}[t]
\includegraphics[height=6cm]{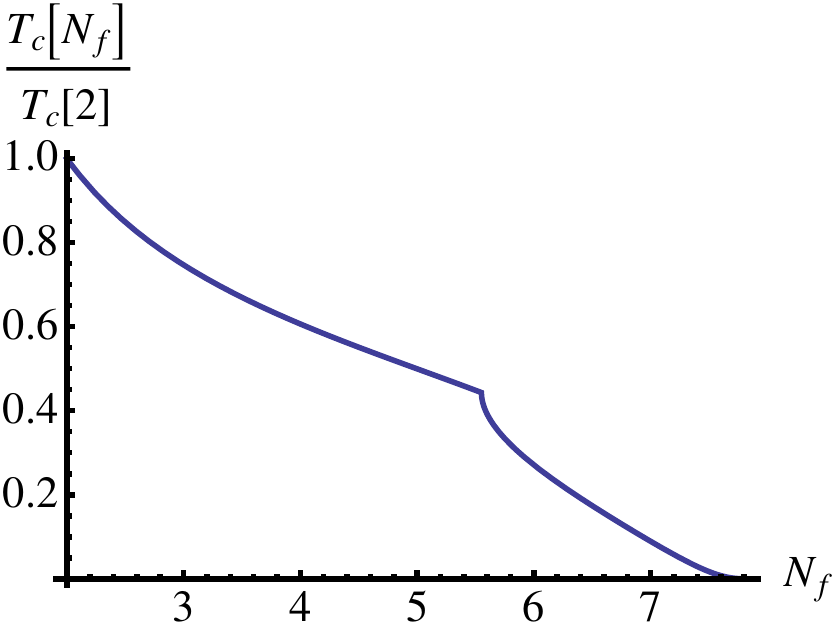}
\includegraphics[height=6cm]{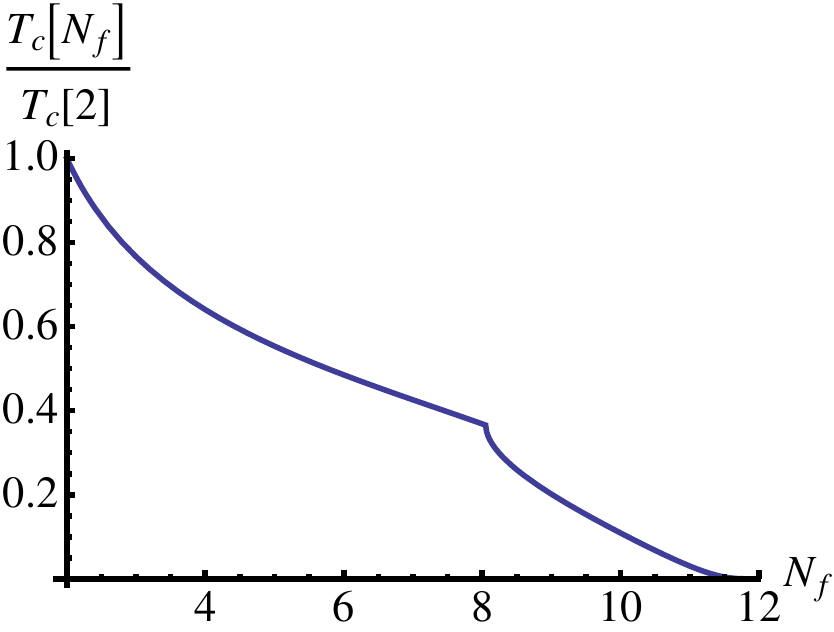}
\caption{The normalized critical temperature as a function of $N_f$ for techniquarks in the fundamental representation of $SU(2)$ (left) and $SU(3)$ right.}\label{twalking}
\end{figure}

The explicit dependence of $\Lambda_c$ on the number of flavors is via the beta function and reads: 
\begin{equation}
\Lambda_c[N_f] = \Lambda_{UV} \exp \left( \int_{0.78 \,\alpha_{c}}^{\alpha_c} \frac{d\alpha}{\beta(\alpha)} \right)\ .
\end{equation}
It is straightforward to see that by approximating the beta function near $\alpha_\ast [N_f]$ leads to:
\begin{equation}
\Lambda_c[N_f]  \approx \Lambda_{UV} \, \left(\frac{\alpha_{\ast}[N_f] - \alpha_c}{\alpha_{\ast}[N_f] - 0.78\, \alpha_c}\right)
^{\frac{1}{\beta_0 \alpha_{\ast}[N_f] }}  \label{BG}\ .
\end{equation}
We can, at this point, plot the normalized walking critical temperature for the two and three colors case for fermions in the fundamental representation, respectively, in the left and right panels of Fig.~\ref{twalking}. 

From the pictures one immediately notices the drop of the temperature near the critical number of flavors above which chiral symmetry restores. For very low number of flavors we observe the geometric dependence on the number of flavors dictated by chiral perturbation theory. Our crude approximation in the intermediate region leads to a discontinue value of the derivative of the walking critical temperature with the respect to the number of flavors at the point when the second coefficient of the beta function vanishes. We expect these two distinct regimes, the drop of the critical temperature at low number of flavors due to straightforward chiral dynamics, and the one near the lower end of the conformal window due to the near conformal dynamics to help disentangling the two regions of the phase diagram of gauge theories of fundamental interactions. Within these simple approximations for QCD we expect the conformal dynamics to dominate above eight flavors while for the two colors case above five flavors \cite{Dietrich:2006cm}. 
 
The prediction of Braun and Gies \cite{Braun:2009ns} on the flavor dependence of the critical temperature is encoded in our approach in \eqref{BG}, however we observe a further exponential suppression related to the dependence of the dynamical mass on the number of flavors in \eqref{sigma}. The main difference is that we have linked the critical temperature to the dynamical mass while in the approach of \cite{Braun:2009ns} the  critical temperature is obtained using dimensional analysis directly from the energy scale in \eqref{BG}.

\section{Conclusions and Impact on Cosmology}
We provided the thermal properties of technicolor models using chiral perturbation theory. At sufficiently large temperatures we explored the melting of the techni-chiral condensate. We used the two-loop expressions at nonzero temperature to estimate the geometric dependence of the critical temperature on the number of flavors for technimatter in complex, real and pseudoreal representations of the underlying gauge group and also in the presence of multiple distinct matter representations. 

By combining our results with the dependence of the pion decay constant on the number of flavors near the lower end of the conformal window, deduced using the Schwinger-Dyson approximation, we have shown that the critical temperature walks, i.e. decreases near the lower bound of the conformal window. We pointed out the existence of two distinct regimes, one at low number of flavors according to which the temperature drop as function of flavors is due to chiral pion dynamics and the one for large number of flavors which is dominated by the near conformal dynamics.

We now show that the results obtained in the previous sections are also relevant when describing the early universe phase transition via models of dynamical electroweak symmetry breaking \cite{Cline:2008hr,Jarvinen:2009mh,Jarvinen:2009wr,Jarvinen:2009pk}. 

In particular the knowledge of the critical temperature, in units of the electroweak scale, plays a major role when estimating the order of the phase transition. More precisely one estimates the order of the phase transition by computing the ratio, of the thermal value of the scalar fermion condensate $\phi_*$ (the composite Higgs field) at the bubble nucleation temperature $T_*$, and $T_*$ itself. A strong first order phase transition requires: $\phi_*/T_* \gtrsim 1$. We first identify the bubble nucleation temperature with the critical temperature deduced in the \ref{rt} section and shown explicitly at one-loop in \eqref{onetemperature}. 

A first order phase transition is expected to occur \cite{Cline:2008hr,Jarvinen:2009mh,Jarvinen:2009pk} for $T_* \lesssim 250$~GeV. The origin of this result can be easily explained: $\phi$ evaluated in the broken phase at the minimum decreases with increasing critical temperature and since at zero temperature $\phi_* \lesssim \phi(T=0) = 246$~GeV the higher is the nucleation (critical) temperature with respect to the electroweak scale the smaller is the ratio $\phi_*/T_*$ with respect to unity.

 Therefore, the phase transition in MWT, NMWT and UMT appears to be either of second order or weak first order. 

One can increase the order of the dynamical electroweak phase transition by considering the many flavors gauged under the electroweak symmetry. In this case from \eqref{onetemperature} it is evident that the critical temperature decreases with the number of flavors favoring stronger first order phase transitions.

Our results provide relevant benchmarks for lattice computations of the conformal window \cite{Catterall:2007yx,Appelquist:2007hu,DelDebbio:2008wb,Shamir:2008pb,Deuzeman:2008sc,DelDebbio:2008zf,Catterall:2008qk,Svetitsky:2008bw,DeGrand:2008dh,Fodor:2008hm,Fodor:2008hn,Deuzeman:2008pf,Deuzeman:2008da,Hietanen:2008vc,Jin:2008rc,DelDebbio:2008tv,DeGrand:2008kx,Fleming:2008gy,Hietanen:2008mr,Appelquist:2009ty,Hietanen:2009az,Deuzeman:2009mh,Fodor:2009nh,DeGrand:2009mt,DeGrand:2009et,Hasenfratz:2009ea,DelDebbio:2009fd,Fodor:2009wk,Fodor:2009ar,Appelquist:2009ka,DeGrand:2009hu,Catterall:2009sb,Bursa:2009we,Lucini:2009an,Pallante:2009hu,Bilgici:2009kh,Machtey:2009wu,Moraitis:2009xt,Kogut:2010cz,Hasenfratz:2010fi,DelDebbio:2010hu,DelDebbio:2010hx} at nonzero temperature  for any nonsupersymmetric vector-like gauge theory with fermionic matter. There has already been much interest in investigations of the thermodynamical properties of the conformal window on the lattice. These studies were pioneered by Deuzeman, Lombardo and Pallante \cite{Deuzeman:2008pf,Deuzeman:2008da} for theories with fermions in the fundamental representation and by Kogut and Sinclair \cite{Kogut:2010cz} for minimal walking models. The temperature dependence on the number of flavors stemming from chiral pertrubation theory is exact while the model dependence is encoded in the flavor dependence of the pion's decay constant.  Our predictions can be also used to verify results obtained using gauge-gravity duality \cite{Jarvinen:2009fe,Alanen:2010tg} and, in the future, to further analyze electro-magnetic gauge duality \cite{Sannino:2009qc,Sannino:2009me,Sannino:2010fh}. 

\acknowledgments
We thank  J. Braun, L. Carloni, D. D. Dietrich, M.T. Frandsen, H. Gies, M.P. Lombardo, E. Pallante and C. Pica for relevant discussions and useful comments. 
 

\end{document}